\documentclass[twocolumn]{webofc}

\usepackage[varg]{txfonts}  
\usepackage{color}
\usepackage{bm}

\newcommand{\md}{\textcolor{black}} 

\begin{document}

\title{Highly strained mixtures of \md{bidimensional} soft and rigid grains: an experimental approach from the local scale }

\author{
\firstname{Jonathan}    \lastname{Bar\'{e}s}              \inst{1} \fnsep\thanks{\email{jb@jonathan-bares.eu}} \and
\firstname{Manuel}      \lastname{C\'{a}rdenas-Barrantes} \inst{1} \and
\firstname{David}       \lastname{Cantor}                 \inst{2} \and
\firstname{\'{E}milien} \lastname{Az\'{e}ma}              \inst{1,3} \and
\firstname{Mathieu}     \lastname{Renouf}                 \inst{1}
}

\institute{
LMGC, Université de Montpellier, CNRS, Montpellier, France \and
Department of Civil, Geological and Mining Engineering, Polytechnique Montr\'{e}al, Qu\'{e}bec, Canada \and
Institut Universitaire de France (IUF), Paris, France
}

\abstract{
Granular systems are not always homogeneous and can be composed of grains with very different mechanical properties. To improve our understanding of the behavior of real granular systems, in this experimental study, we compress 2D bidisperse \md{systems} made of both soft and rigid grains. By means of a recently developed experimental set-up, \md{from the measurement of the displacement field we can} follow all the mechanical observables of this granular medium from the inside of each particle up-to the whole system scale. We \md{are able to} detect the jamming transition from these observables and study their evolution deep in the jammed state for packing fractions as high as $0.915$. We show the uniqueness of the behavior of such a system, \md{in which way} it is similar to purely soft or rigid systems and how it is different from them. This study constitutes the first step toward a better understanding of mechanical behavior of granular materials that are polydisperse in terms of grain rheology.
}

\maketitle

\section{Introduction} \label{intro}

Loaded granular materials made of highly deformable particles are ubiquitous in nature and industry from cell monolayer growth \cite{angelini2011_pnas,martin2004_dev} to foam and emulsion compression \cite{bolton1990_prl,katgert2010_epl,brujic2007_prl} and even metal or plastic powder sintering \cite{kim1987_ijp,bares2014_prl}. In some cases, because of impurities or for technical and recycling purposes, rigid, undeformable particles are mixed among soft ones \cite{tsoi2011_geo,khatami2020_gi}. This is the case for example in sand-rubber granular layers used as low-cost seismic \md{isolations} \cite{tsiavos2019_sdee}. 

These kinds of mixed soft-rigid granular systems display singular behaviors different from purely rigid or \md{purely} soft systems. Among other purposes, modifying the mixing ratio and the particle nature permits to tune the mechanical properties of a granular medium. In order to optimize their industrial applications and to control \md{their} behavior in \md{natural phenomena}, it is needed to know the connection between their macroscopic behaviors and their microscale composition and organisation. Because granular matter is naturally \md{multiscale}, \md{a} clear understanding of its global behavior necessitates to figure out and to measure what happens at the local/grain scale and at all the intermediate scales up-to the global one. Due to the already complex process of measuring local mechanical properties of each grain for fewly deformed particles \cite{zadeh2019_gm} the challenge is even more important of measuring them for soft highly deformed particles. 

Very recently this challenge has been overcome \cite{vu2019_pre,vu2019_pre_2} using an experimental method coupling Digital Image Correlation (DIC) and very \md{acurate} imaging. In this paper, we extend this method to the so called case of mixtures of soft and rigid particles. We show how this experimental set-up is \md{modified} to compress \md{these} mixtures and how the image post-processing is modified accordingly. A selection of resulting experimental data from the inside of the particle up-to the system size is presented. This allows us to understand better the mechanical behavior of these materials and will permit to improve their compression models \cite{cardenas2020_pre}.

\section{Experimental method} \label{method}

\begin{figure}[h]
\centering
\includegraphics[width=0.99\columnwidth]{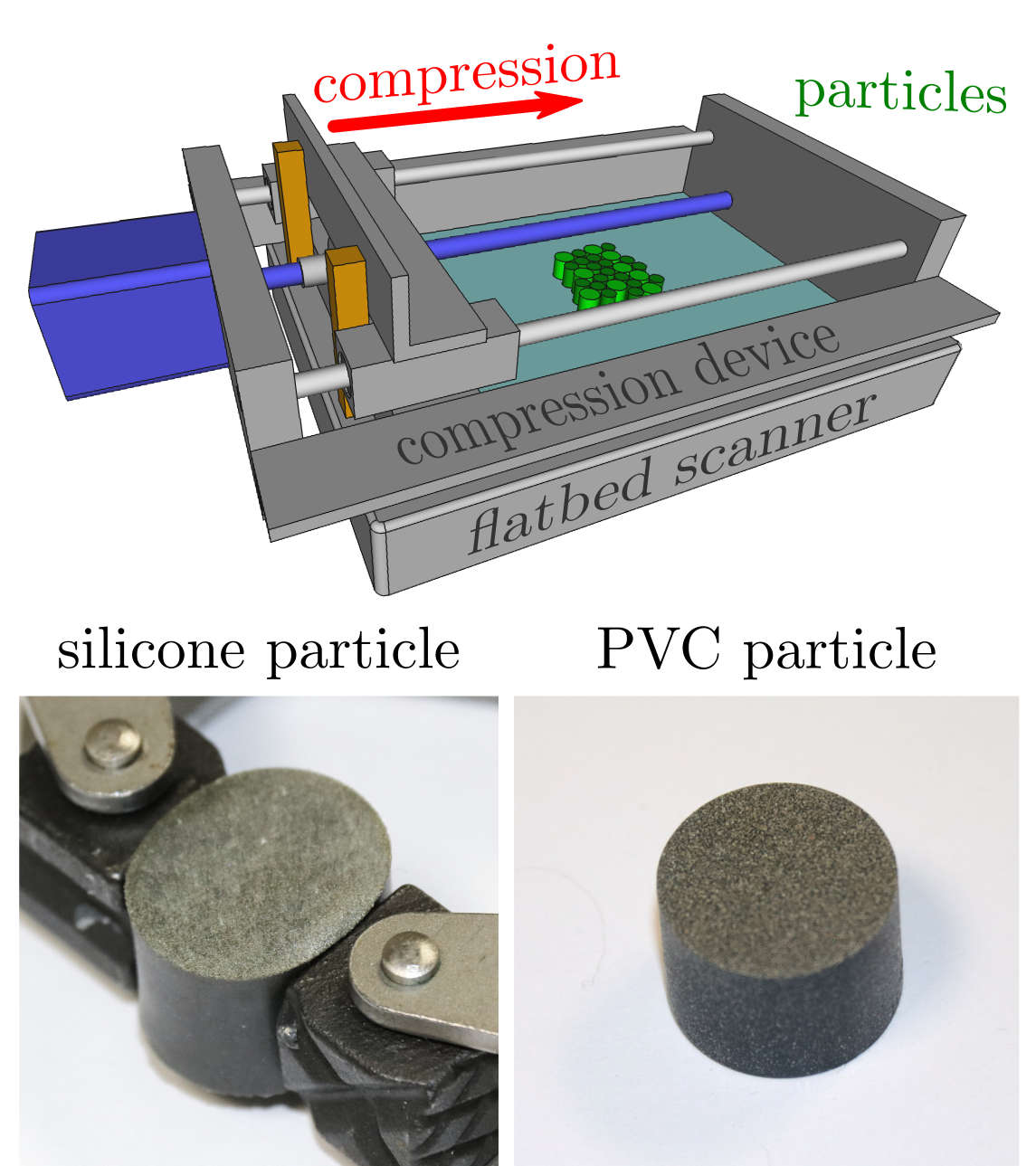}
\caption{Experimental set-up. Top: 3D \md{drawing} of the compression device and particle packing laying on a photographic flatbed scanner \cite{scanner}. Bottom-left: squeezed soft silicone particle \cite{silicone,vu2019_pre_2}. The top shiny aspect is due to a layer of thin metallic glitter \cite{silver_glitter}. Bottom-right: rigid PVC particle. The top shiny aspect is due to silver paint. \md{During experiments particles are used with their shiny part in contact with the scanner.}}
\label{fig_1}
\end{figure}

In Fig. \ref{fig_1} we show the experimental set-up already presented in \cite{vu2019_pre_2}. It is composed of a bidimensional piston of initial dimension $270\times202$ mm$^2$ \md{which permits to compress uniaxially the system}. A stepper motor rotates a screw which \md{translate the moving edge of the piston inward in an accurate manner ($25$ $\mu$m per motor step)}. Two force sensors are attached to this latter, they record the global stress evolution ($\sigma$) when compressing a mixed granular system. The global stress is measured continuously with a frequency of $100$ Hz while the system is compressed stepwisely. At each step the piston moves of $0.5$ mm at a speed of $2$ mm/min to make \md{it} sure to stay in the quasistatic regime. Then, the system rests during $1$ min to relax and is imaged from below with a flatbed \cite{scanner} with a resolution of $2400$ dpi ($10.6$ $\mu$m/px).

The granular \md{systems are} composed of soft hyperelastic cylinders made of silicone \cite{silicone,vu2019_pre_2} and of rigid \md{cylinders} made of PVC (see Fig. \ref{fig_1}). Their Young moduli are $0.45$ MPa \cite{vu2019_em} and $1.2$ GPa, respectively. Both soft and rigid granular packings are bidisperse cylinders of diameters $20$ mm and $30$ mm and of height $15$ mm. For the \md{experiments} presented in this paper, $96$ particles are used, $1/2$ of them being soft and $2/3$ of them being small. This last ratio is equally spread among soft and rigid particles, this avoids phase crystallization. \md{The ratio of soft to rigid particles, $\kappa = \rm{n}_{\rm{soft}}/\rm{n}_{\rm{total}}$ is varied from $0.2$ to $0.8$.} To avoid basal friction the scanner glass is coated with oil which also lubricate the contacts making particles almost frictionless. \md{Namely the measured friction coefficient is $0.05 \pm 0.02$.}

To measure displacement fields on the particle bottom faces, the DIC method requires a thin random pattern with a high optical contrast. In the case of soft particles, this is obtained by coating the silicone with a very thin layer of \md{shiny} glitter \cite{silver_glitter}. For rigid particles, PVC is simply coated with \md{silver metallic} paint. In both cases, the glitter characteristic size is around $25$ $\mu$m. \md{This} creates a random pattern with a correlation length of \md{about} $50$ $\mu$m on the bottom face of each particle.

\md{For each experiment a set of about $100$} pictures is obtained with this experimental set-up by compressing the system over $15$\%. Pictures are then post-processed with an algorithm modified from \cite{vu2019_pre_2, DIC_code}. First, particles are detected individually by thresholding the initial/undeformed picture. Each particule is then tracked in the full set of pictures by using image correlation. Then, from the measurement of the particle rotation and translation, subsets of images are extracted following each particle and correcting its solid rigid motion. For soft particles a DIC algorithm aimed for large \md{deformations} already presented in \cite{vu2019_em} is used to obtain the displacement field ($\bm{u}$) inside each particle. For the rigid particles a more classical DIC algorithm is used. This latter correlates all images with the initial one as classically done for systems \md{in} the small deformation assumption. In both cases the correlation cell size is self-adjusted to take random pattern inhomogeneities into account as already presented in \cite{vu2019_pre_2}. 

Then, for soft particles the deformation gradient tensor is computed from $\bm{u}$ \cite{taber2004_book}:
\begin{equation}
	\bm{\bm{F}}=\bm{\nabla}\bm{u}+\bm{\bm{I}}
\end{equation}
with $\bm{\bm{I}}$ being  the  second  order  identity  tensor. The right Cauchy-Green strain tensor is finally derived \cite{taber2004_book}:
\begin{equation}
	\bm{\bm{C}}=\bm{\bm{F}}^T\bm{\bm{F}}
\end{equation}
For rigid particles, the infinitesimal strain tensor is just directly computed ($\bm{\bm{\varepsilon}}=1/2((\bm{\nabla}\bm{u})^T+\bm{\nabla}\bm{u})$). From $\bm{u}$ the evolution of the exact particle boundaries is also deduced which permits to get all geometric \md{observables} for both particles and voids, but also contacts, at each compression step. \md{Contacts are detected by considering first the proximity of each point of the particles boundaries with the boundaries of neighbor particles. Then, for each point closer than a certain threshold the von Mises strain around this point is measured and the contact is assumed to be real if this value is above a certain threshold as detailed in \cite{vu2019_pre_2}.} For the sake of conciseness only a meaningful selection of \md{the different} mechanical and geometrical observables are presented here.

\section{Results and discussions} \label{result}

\begin{figure}[h]
\centering
\includegraphics[width=0.99\columnwidth]{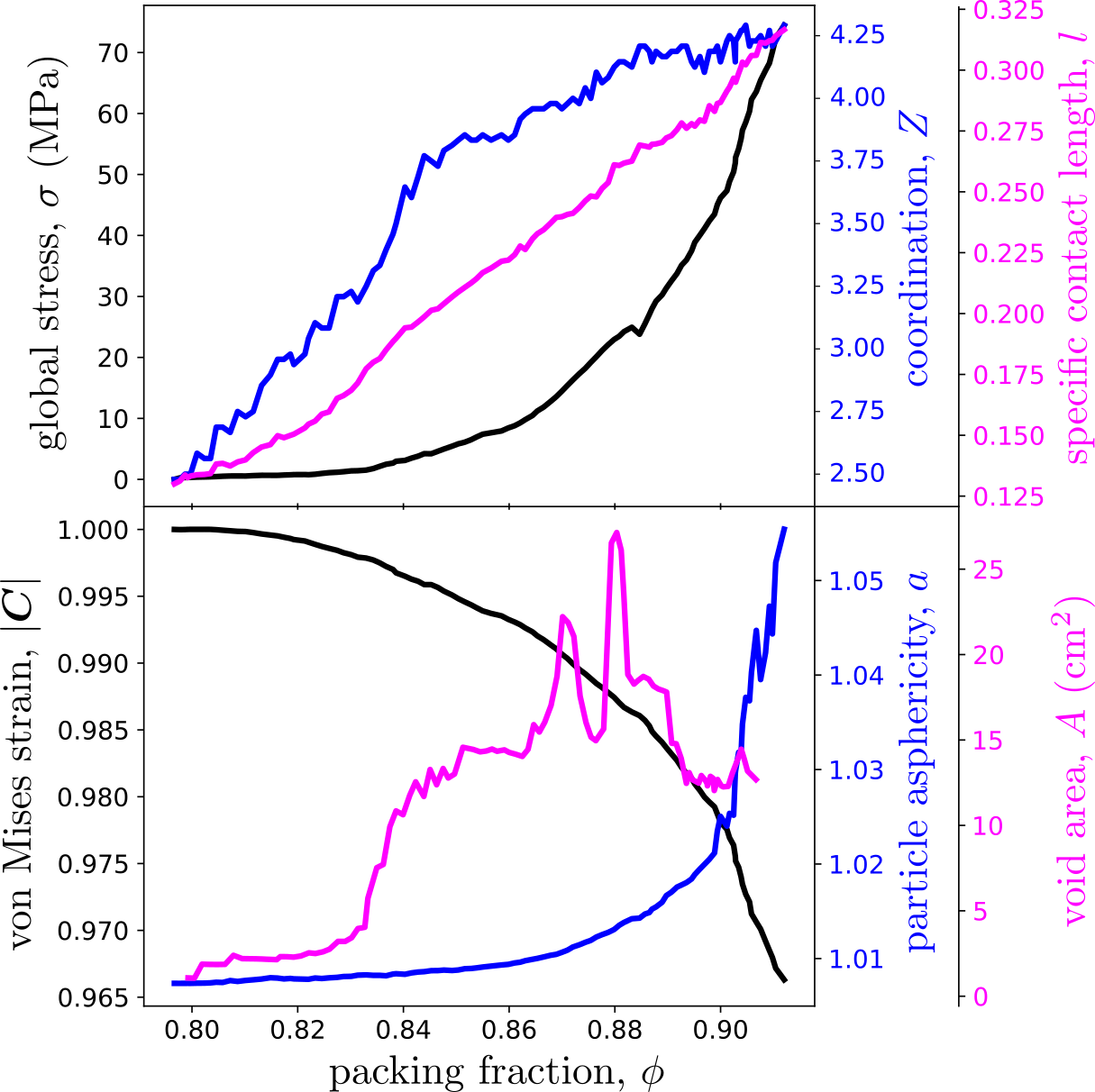}
\caption{Global observables. Variation of the global stress ($\sigma$), coordination ($Z$), specific contact length ($l$), average value of the von Mises of the right Cauchy-Green strain tensor ($|\bm{\bm{C}}|$) of soft particles, particle asphericity ($a$) and void area ($A$) as a function of the packing fraction ($\phi$) for a compression experiment made with $50$\% of rigid particles ($\kappa = 0.5$).}
\label{fig_2}
\end{figure}

As presented in Fig. \ref{fig_2}, \md{a granular system with $\kappa = 0.5$ is compressed for a packing fraction ($\phi$) going from $0.795$ to $0.915$}. While loading, the global stress first stays close to $0$ MPa and then increases progressively from the jamming point \cite{liu1998_nat}. This occurs for $\phi$ close to \md{$\phi_j = 0.84$}, as already observed in many other experimental studies \cite{behringer2018_rpp}. It also corresponds with a coordination value slightly above $3$ as predicted theoretically \cite{van2009_jpcm} since particles are slightly frictional. In Fig. \ref{fig_2} we also plot the evolution of the specific contact length, which is the average of \md{the ratio of the particle perimeter in contact with another particle or with the system boundaries ($l_{\rm contact}$) over the particle perimeter: $l = \left< l_{\rm contact}/\rm{perimeter}\right>_{\rm particles}$.} This quantity increases linearly with $\phi$, which is not in agreement with the Hertz's contact law in 2D \cite{johnson_book}. This last point is not surprising since soft particles are far from the small deformation assumption \md{in our study}. 

Figure \ref{fig_2} also shows that the average value of the von Mises strain ($|\bm{\bm{C}}|$) in soft particles decreases parabolically all along the compression while the average soft particle asphericity first gently increases and then sharply rises after $\phi=0.88$. This quantity is computed as the ratio between the actual particle area and the area deduced from the perimeter assuming the particle would be circular. The packing fraction $\phi=0.88$ also corresponds with the maximum value of the total void area which first increases, since void clusters form, and then decreases, since particles are squeezed and reduce these \md{cluster} areas, as already observed for fully soft systems \cite{vu2019_pre_2}.

\begin{figure}[h]
\centering
\includegraphics[width=0.99\columnwidth]{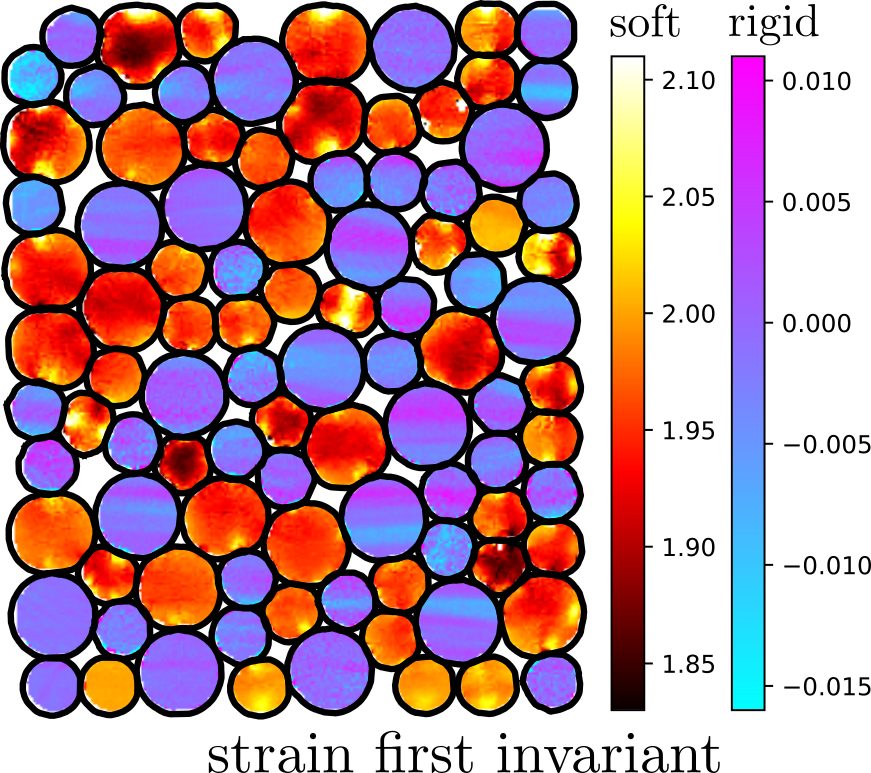}
\caption{Local observables. Field of the first invariant of the right Cauchy-Green strain tensor ($\bm{\bm{C}}$) for soft particles and of the infinitesimal strain tensor ($\bm{\bm{\varepsilon}}$) for rigid particles. These fields are computed for a compression experiment made with $50$\% of rigid particles at a packing fraction $\phi = 0.89$ far above the jamming point.}
\label{fig_3}
\end{figure}

In Fig. \ref{fig_3}, the first invariant of the strain fields measured in each particle is shown for $\phi = 0.89$. For soft particles the right Cauchy-Green strain tensor ($\bm{\bm{C}}$) is used while it is the infinitesimal strain tensor ($\bm{\bm{\varepsilon}}$) \md{which is used} for rigid particles. The field for soft particles distinctly shows high strain chains with clear areas near the contact points where the deformation is maximum. On the contrary the patterns observed on the rigid particles suggest that the imaging method is not accurate enough to \md{highlight} grain deformations. This is not surprising since the level of compression is rather low and the \md{ratio of soft to rigid} particle stiffness covers several orders of magnitude: \md{deformation is almost exclusively carried by soft particles.} The lines that appear on the rigid particles actually correspond with the small speed inhomogeneities of the \md{imaging method}.
 
\begin{figure}[h]
\centering
\includegraphics[width=0.99\columnwidth]{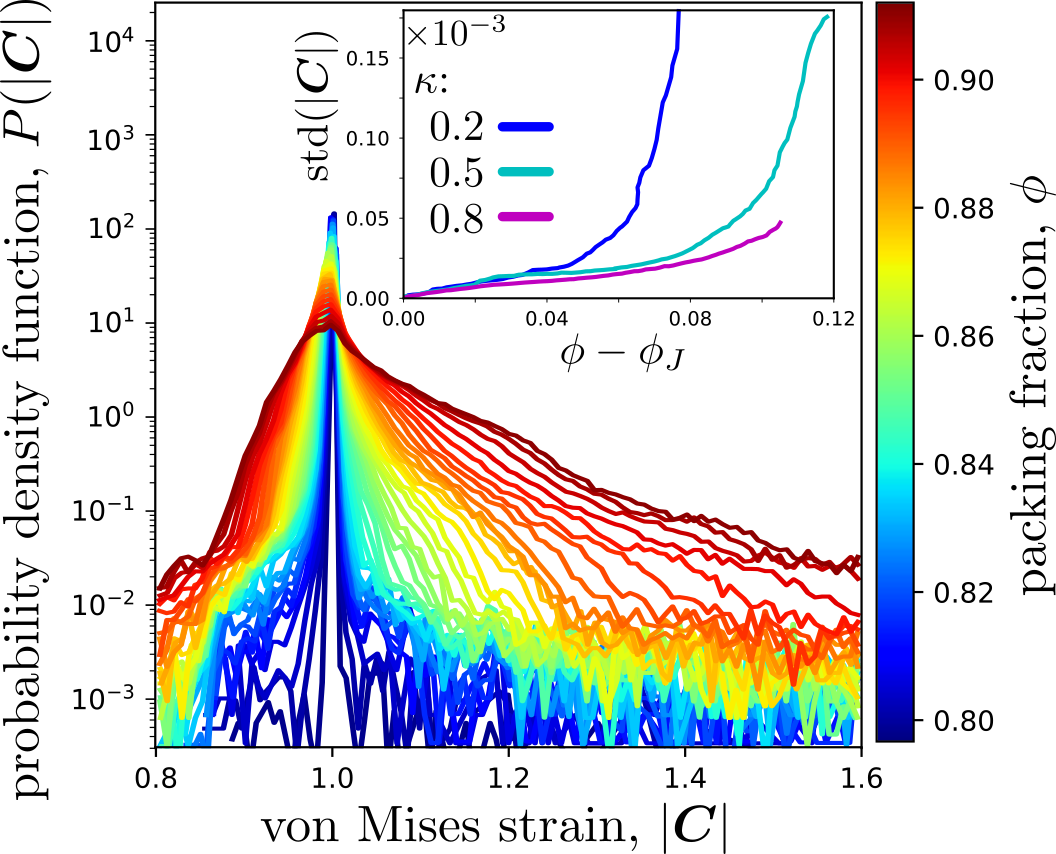}
\caption{Statistical observation. \md{Main panel}: Evolution of the PDF of the value of the von Mises of the right Cauchy-Green strain tensor ($\bm{C}$) measured in soft particles as a function of the packing fraction for a compression experiment made with $50$\% of rigid particles. Scale is semi-log. \md{Inset: Evolution of standard deviation of $\bm{C}$ for different particle mixes ($\kappa$) as function of the distance to the jamming point.}}
\label{fig_4}
\end{figure}

In Fig. \ref{fig_4}, we present the evolution of the strain field statistics. More exactly, we plot the probability density functions (PDF) of the von Mises values of the right Cauchy-Green strain tensor ($|\bm{\bm{C}}|$) in soft particles for different packing fractions. For gently squeezed states, the PDF is a Gaussian centered around $1$ which corresponds with the fact that the particles are mainly undeformed. While the system is loaded, the PDF gets broader and broader and the average value slightly shifts to the left which corresponds with the global compression of the system. For densely packed states, the PDFs display an exponential tail with a coefficient decreasing with the packing fraction. This exponential tail is reminiscent of what is observed for force values in packing of rigid granular systems \cite{radjai1996_prl} and remains unchanged from observations made on purely soft systems \cite{vu2019_pre_2}. \md{In the inset of Fig. \ref{fig_4}, the evolution of the standard deviation of $\bm{C}$ is shown for different ratios between soft and rigid particles. When the system is mainly composed of rigid particles ($\kappa = 0.2$), this value rapidly increases right after the jamming point. In this case lot of particles undergo large deformation and display more extreme strain values. On the contrary, when the system is composed of more soft particles ($\kappa = 0.8$) the deformation is shared over more particles and less extreme strain values are observed so the standard deviation increases less rapidly.}

\section{Conclusion} \label{conclusion}

In summary, we have developed an experimental tool that \md{can} permit to measure all mechanical and geometrical observables from \md{the displacement field at} the particle scale for bidimensional granular systems under uniaxial compression, whatever the particle mechanical properties. \md{We believe this measurement method at the \textit{sub-micro scale} will permits to model granular matter in a more efficient way to predict its behavior.} It is applied here to the case of a mixed granular \md{media} made of soft squeezable particles and rigid ones. The system is loaded step by step and imaged very accurately, while \md{the} global force applied to it is recorded continuously. By mean of DIC, the evolution of mechanical fields inside each particle are measured which permits to deduce observables as diverse as packing fraction, coordination, contact length, particle asphericity, void area and strain fields to name a few.

More specifically we have shown that our system crosses the jamming transition for a packing fraction and a coordination value close from what has been observed in the case of purely rigid systems. This point validate our experimental approach. \md{We have also shown that the average von Mises strain in soft particles decreases with the exact same tendency as the global stress increases in the system. Since rigid particle deformation is negligible, this suggests that rigid particles do not carry more load than the soft ones. The loading seems to be balanced between soft and rigid particles.} The stress vs. packing fraction curve is also similar to what has been observed in a recent numerical study \cite{cardenas2020_pre}. 

Deep in the jammed state the PDF of the von Mises strain tensor displays an exponential tail as already observed for purely soft systems and similarly to what is observed for the \md{intergranular} force PDF \md{of} purely rigid systems. \md{Also, the standard deviation of this quantity increases more rapidly for systems composed of more rigid particles. This is explained by the fact that deformation is mainly carried by soft particles and that they are individually more deformed when they are less for a given deformation level.}

Beyond the study of mixed soft-rigid granular systems, we believe that this experimental tool offers an opportunity to catch experimentally the behavior of breakable and polydisperse (in size, shape and material) \md{particles} under different loading geometries. 

{\small
Gille Camp, St\'{e}phan Devic and R\'{e}my Mozul are greatly thanked for their technical support. We warmly thank Vincent Huon for for fruitful discussions.
}
\newline
\bibliographystyle{unsrt}
\bibliography{biblio}

\end{document}